# Learning Conventions in Multiagent Stochastic Domains using Likelihood Estimates

**Craig Boutilier**
Dept. of Computer Science
University of British Columbia
Vancouver, BC V6T 1Z4
*cebly@cs.ubc.ca*

## Abstract

*Fully cooperative* multiagent systems—those in which agents share a joint utility model—is of special interest in AI. A key problem is that of ensuring that the actions of individual agents are coordinated, especially in settings where the agents are autonomous decision makers. We investigate approaches to learning coordinated strategies in stochastic domains where an agent's actions are not directly observable by others. Much recent work in game theory has adopted a Bayesian learning perspective to the more general problem of equilibrium selection, but tends to assume that actions can be observed. We discuss the special problems that arise when actions are not observable, including effects on rates of convergence, and the effect of action failure probabilities and asymmetries. We also use likelihood estimates as a means of generalizing *fictitious play* learning models in our setting. Finally, we propose the use of maximum likelihood as a means of removing strategies from consideration, with the aim of convergence to a *conventional* equilibrium, at which point learning and deliberation can cease.

## 1 Introduction

The design of systems of multiple autonomous agents that interact in various ways (pursuing their own ends or compatible goals) has attracted a great deal of attention in AI. Of special interest are systems in which individual agents share the same goals or utility function—in such *fully cooperative* settings, the agents collectively act toward common desired ends. While more general problems involving the interaction of potentially self-interested agents have received the bulk of attention in distributed AI, fully cooperative problems naturally arise in task distribution. For example, a user might assign some number of autonomous mobile robots, or perhaps software agents, to some task, all of which should share the same utility function (namely, that of the user). For certain purposes, it may make sense to model a business or organization in a similar way.

A key difficulty in cooperative multiagent systems is that of ensuring that the actions of individual agents are coordinated so that the shared goals are achieved efficiently. This is especially important in settings where the agents are autonomous decision makers (and preprogrammed coordinated strategies are not available), as in the situations mentioned above. One natural way to view the coordination problem is as a $n$-person cooperative game. From the perspective of game theory, we are interested in $n$-person games in which the players have a shared or *joint* utility function; that is, any outcome of the game has equal value for all players.

In this paper, we study aspects of the coordination problem from the perspective of $n$-player repeated games. A set of agents find themselves in a situation which requires coordinated action (viewed as a single-stage decision problem), but can encounter this situation repeatedly.[1] Methods such as allowing agents to communicate their intentions beforehand or imposing specific behaviors (e.g., by means of a central controller or the use of social laws) may ensure that agents behave in a coordinated fashion. However, our interest in this paper is in methods that enable agents to *learn* their component of a coordinated policy through repeated experience with the game situation.

Learning techniques have been well-studied in game theory, not only for coordination in cooperative games, but also for the more general problem of *equilibrium selection* [12, 5]. Models applied to this problem include *fictitious play* [13] and *Bayesian best-response* methods [8, 19, 4] (evolutionary models have also attracted a great deal of attention [1, 11]). These have especially nice behavior in coordination problems [19]. However, these models tend to assume that each agent can observe the exact action performed by all others at each interaction. Such *action observable* scenar-ios will likely be rare in practice, especially when individual actions have stochastic effects. Even if states of the system (and thus action outcomes) are fully observable—as they might be in a Markov decision model—it is unlikely that agents will have access to the actual action another agent *attempted* (and hence the "intentions" of the other agent).

---

[1]This scenario is appealing in its simplicity, but is not an overly realistic picture of multiagent decision problems. However, our interest in repeated single-stage games is motivated by a decomposition of sequential cooperative problems (see below).

We focus our attention on games where actions are stochastic, and actions are not directly observable. In general, agents can observe only the state resulting from the actions of the group of players; but they can use this observation to assess the probability that other agents performed particular actions. The introduction of this type of uncertainty and partial observability is rather simple to model, but it has some rather surprising effects on convergence to coordinated action in the Bayesian best-response model, which we examine here. We also adapt fictitious play to this unobservable action setting through the use of likelihood estimates, and show that convergence is generally much better than in the Bayesian model.

Finally, we consider the problem of learning *conventions* [9, 16]. One difficulty with stochastic games and models that require constant learning is that a run of "bad luck" can force agents out of a coordinated equilibrium. More serious are the computational implications of constantly updating beliefs and computing a best response for every interaction. Following Lewis [9], we take an interest in conventional behavior. Agents should converge to a common understanding and, once realizing that they have reached a coordinated equilibrium, should never be forced to reconsider how to act. Of course, care must be taken to ensure this understanding is based on common knowledge, or globally accessible information. To this end, we propose the use of "globally accessible" likelihood estimates to rule out particular ways of acting, until only a conventional method of acting remains whenever possible.

We describe the basic framework of coordination games in Section 2, as well as their application to multiagent sequential decision processes (in the form of multiagent Markov decision processes). In Section 3 we detail classic models from game theory for learning coordinated actions, in particular fictitious play and Bayesian methods. We also point out the difficulty asymmetric coordination games pose for such methods. In Section 4, we extend these models in rather obvious ways to deal with stochastic, partially-observable actions. We study a number of properties of these models and how convergence is affected by them. We address the problem of convention in Section 5, proposing an extension of fictitious play dynamics whereby likelihood estimates for optimal joint actions are used to rule out possible courses of action.

Experimental results are provided to illustrate the performance of these methods. We focus (primarily, not exclusively) on $2 \times 2$ games to keep the exposition clear; but most of the conclusions we draw can be applied more broadly.

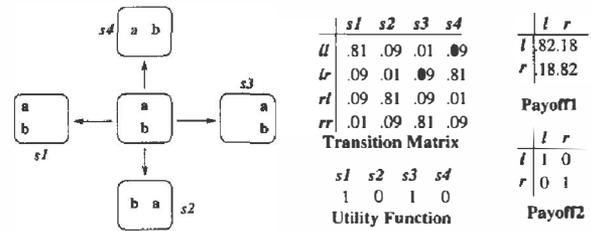

Figure 1: A Two-Agent Coordination Problem

## 2 Coordination Games

### 2.1 Single Stage Games

We take as the basic object of study an $n$-*player cooperative state game*. We assume a collection $\alpha$ of $n$ (heterogeneous) agents, each agent $i \in \alpha$ having available to it a finite set of *individual actions* $A_i$. The game takes place at a given state, at which each agent chooses (independently) an action to perform. The chosen actions collectively constitute a *joint action*, the set of which is denoted $A = \times_{i \in \alpha} A_i$. The game also has a set of *outcomes states* $S$: each joint action causes a transition to some outcome state $s \in S$ according to a fixed distribution. We use $Pr^a(s)$ to denote the probability of outcome $s$ when the joint action $a$ has been executed. Finally, we associate a utility $U(s)$ with each $s \in S$. Intuitively, each agent receives reward $U(s)$ if the joint action adopted by the agents results in $s$; the game is thus fully cooperative, for agents cannot do better by making things worse for others.[2]

We note that state games are essentially single-stage *extensive form* games; but it is convenient to sometimes convert them to the corresponding *strategic form* (or their *normal representation*) [12]. Each joint action $a$ can be associated with its expected utility, $\sum_{s \in S} Pr^a(s) \cdot U(s)$, and states can be done away with, resulting in a strategic form game. However, the existence of distinct outcome states is crucial in the learning models we adopt below. In particular, the states provide indirect information about action choices in cases where actions are not directly observable. Conversion to strategic form precludes the use of this partial information; however, when actions are perfectly observable, we will often use strategic form.

As an example, consider the $2 \times 2$ game illustrated in Figure 1, in which two agents, $A$ and $B$, can move left ($l$) or right ($r$) (say, toward a particular goal). The agents are rewarded with utility 1 if they both end up in the same location—either both left ($s1$) or both right ($s3$)—and utility 0 otherwise. The actions available to the agents are stochastic, so that if $A$ executes action $l$, it will end up in the left location with probability 0.9 and in the opposite lo-cation with probability 0.1. This results in the transition matrix shown. Should we convert this game to strategic form, the Payoff Matrix 1 describes the expected utility of the given joint actions. We will also have occasion to use the deterministic version of this game, where each joint action has the obvious outcome: Payoff Matrix 2 characterizes this game. We note that in the deterministic game, an agent observing the outcome state is equivalent to observing its companion's action directly.

---

[2] A general $n$-person state game simply requires that $U$ take agents as arguments as well as states to allow for competition; i.e., $U(s, i)$ denotes the utility of state $s$ to agent $i$.

Given such a game, we want the agents to discover an optimal course of action. Unfortunately, the optimal action for each agent generally depends on the choices of other agents. The typical solution concept adopted in game theory, that of a *Nash equilibrium*, allows us to break out of potential circularities as follows.

A *randomized strategy* for agent $i$ at state game $G$ is a probability distribution $\pi \in \Delta(A_i)$ (where $\Delta(A_i)$ is the set of distributions over the agent's action set $A_i$). Intuitively, $\pi(a^i)$ denotes the probability of agent $i$ selecting the individual action $a^i$ when playing the game. A strategy $\pi$ is *deterministic* if $\pi(a^i) = 1$ for some $a^i \in A_i$.

A *strategy profile* for $G$ is a collection $\Pi = \{\pi_i : i \in \alpha\}$ of strategies for each agent $i$. The expected value of acting according to a fixed profile can easily be determined. We note that if each strategy in $\Pi$ is deterministic, we can think of $\Pi$ as a joint action, since each agent's action is fixed. A *reduced profile for agent* $i$ is a strategy profile for all agents but $i$ (denoted $\Pi_{-i}$). Given a profile $\Pi_{-i}$, a strategy $\pi_i$ is a *best response* for agent $i$ if the expected value of the strategy profile $\Pi_{-i} \cup \{\pi_i\}$ is maximal for agent $i$; that is, agent $i$ could not do better using any other strategy $\pi_i'$.

Finally, we say that the strategy profile $\Pi$ is a *Nash equilibrium* iff $\pi_i \in \Pi$ is a best response to $\Pi_{-i}$, for every agent $i$. In other words, the agents are in equilibrium if no agent could expect a better outcome by unilaterally deviating from its strategy. In general, the interests of different agents can conflict, making equilibrium determination quite difficult. However, in fully cooperative games each agent expects the same reward and can easily determine an interesting set of equilibrium profiles as follows. We first convert the state game to strategic form (by taking expectation of outcome utilities). Any joint action whose expected value is maximal is a (deterministic) Nash equilibrium. Such an equilibrium is called an *optimal joint action* (OJA). If the agents coordinate their choices so that they select an OJA, they are behaving as well as possible.

To illustrate with our example problem (in either the deterministic or nondeterministic version), we see that the OJAs are $\langle l, l \rangle$ and $\langle r, r \rangle$. These strategy profiles offer maximal expected utility for both agents. We note however that being in equilibrium does not guarantee the agents are behaving optimally (in a joint sense). The profile in which each agent adopts a randomized strategy that selects $l$ and $r$ with equal probability is also an equilibrium: given that agent $A$ chooses $l$ or $r$, each with probability 0.5, $B$ has no incentive to change its strategy (similarly for $A$). But this randomized equilibrium is suboptimal, for its expected value is half that of the optimal equilibria.

Nash equilibria, unfortunately, do not solve the coordination problem. While the agents can determine the OJAs quite readily, the problem remains: how do they decide which OJA to adopt? In its most general form, this is precisely the problem of *equilibrium selection* studied in game theory [12, 5]. We take the coordination problem to be that of ensuring agents select individual actions that together constitute an optimal or *coordinated equilibrium*, or OJA.[3]

### 2.2 Multiagent MDPs

While our focus is on simple repeated state games, our motivation is not primarily the solution of repeated, single-stage decision problems. In [2], we propose *multiagent Markov decision processes* (MMDPs) as a framework in which to study multiagent cooperative planning (in decision theoretic contexts). Roughly, MMDPs are Markov decision processes [7, 14, 3] in which actions at each stage are comprised of distinct components performed by individual agents. The aim there is not to coordinate single state strategies *per se*, but to construct coordinated policies for ongoing behavior in different states.

Producing coordinated policies is difficult computationally; but one can gain considerable leverage by decomposing the problem into distinct state games of the type described above, with one game for each state (of a certain type) of the MDP. The coordination problem is then reduced to that of coordinating locally at each of these state games. In [2] we assume that agents can compute the value of coordinated (ongoing) policies at individual states.[4] These "long term" values are used as the outcome utilities in the individual state games. If the agents are able to coordinate locally at each of the state games defined in this way, we can guarantee that they will implement an optimal (sequential) policy [2]. Of course, in an MDP of sufficient horizon, agents will repeatedly encounter the same (or similar) states. For this reason, coordination at single-stage state games has an important application to multistage (especially "process oriented") decision problems.

## 3 Learning with Observable Actions

Solutions to the coordination problem can be divided into three general classes, those based on communication, those based on convention and those based on learning. For example, agents might communicate in order to determine task allocation [18, 17] or simply inform one another what actions they will choose. Conventions (or social laws) might be imposed by the system designer so that optimal joint action is assured [9, 16]—intuitively, a convention restricts (or forces) consideration to a subset of feasible or optimal joint actions (such as the convention of driving on the right hand side of the street). Finally, coordinated action choice might be learned through repeated play of the game, either with the same agents [4, 8, 10] or a random selection of similar agents [1, 15, 11, 19].

---
[3] We note that optimal equilibria need not be deterministic OJAs. E.g., if $A$ had another action $m$ that behaved similarly to $r$, then it could randomize between $m$ and $r$; and if $B$ adopted $r$, an optimal equilibrium would result. However, we will continue to speak as if optimal equilibria are OJAs.

[4] In other words, they can compute the *value function* of the Markov decision process [7, 14]. We refer to [2] for a discussion of the details, benefits and computational implications of this assumption.

We focus here on learning models in which agents repeatedly interact with the same set of players in state games. In this section, we assume that each agent can observe the actions of the others at each interaction. Intuitively, each agent uses its past history to form an estimate of strategies used by the other agents. At each interaction, or play of the game, an agent will choose a best response action to execute, given its predictions (or beliefs) about the behavior of the other agents. Once the game is played, the agent can observe the actual actions chosen by the other players and update its beliefs regarding future play accordingly.

### 3.1 Fictitious Play

One of the simplest learning models for repeated games is *fictitious play* [13]. Each agent $i$ keeps a count $C_{a^j}^j$, $j \in \alpha, a^j \in A_j$, of the number of times agent $j$ has used action $a^j$ in the past. When the game is encountered, $i$ treats the relative frequencies of each of $j$'s moves as indicative of $j$'s current (randomized) strategy. That is, for each agent $j$, $i$ assumes $j$ plays action $a^j \in A_j$ with probability $C_{a^j}^j/(\sum_{b^j \in A_j} C_{b^j}^j)$. This set of strategies forms a reduced profile $\Pi_{-i}$, for which agent $i$ adopts a best response. After the play, $i$ updates its counts appropriately, given the actions used by the other agents.

This very simple adaptive strategy is not guaranteed to converge to equilibrium in general, but will converge for two-person zero-sum games [13]. More importantly, the methods of Young [19] can be applied to our simple cooperative games to show that it is guaranteed to converge to a coordinated equilibrium (that is, the probability of coordinated equilibrium after $k$ interactions can be made arbitrarily high by increasing $k$ sufficiently). We simply require that an agent randomize between all pure best responses when more than one is available.[5] It is also not hard to see that once the agents reach an equilibrium, they will remain there—each best response simply reinforces the beliefs of the other agents that the coordinated equilibrium remains in force. We do not discuss rates of convergence or experiments, since the model is similar to the particular Bayesian methods we describe next.

---

[5] We also require that utilities be rational so that the opportunity to randomize arises (see below).

### 3.2 Bayesian Best-Response Model

A popular method for learning to select equilibria assumes that agents have a prior beliefs, in the form of a probability distribution, over the possible strategies of other agents, use Bayesian update this adjust their priors as experience dictates, and adopt a best response at each interaction based on their current beliefs [8, 4]. In repeated games, one could (and should) technically have priors over another agent's *sequential strategy*, including how it might react to one's current moves in the future [8]. However, the practical difficulties of specifying anything but the simplest priors is evident; this also runs contrary to the spirit of decomposing sequential problems into states games (Section 2.2). So we restrict attention to beliefs about single-stage strategies for the state game $G$.

We assume each agent $i$ has an prior distribution over the strategies that could be adopted by other agents. The *beliefs* of agent $i$ about agent $j$ are represented by a probability distribution over the set of (randomized) strategies $\Delta(A_j)$ agent $j$ might adopt. We denote by $Bel_i(j, \pi_j, s)$ the degree of belief agent $i$ has that $j$ will perform strategy $\pi_j$.

As a general rule, any reasonable prior could be used (provided it does not rule out the choice of some action in the state game). However, we will consider only the case where each agent uses a simple prior, the *Dirichlet distribution*. This can be represented with a small number of parameters and can be updated and used quite easily. Let $n$ be the cardinality of $j$'s action set. Agent $i$'s beliefs about $j$ are represented by the Dirichlet parameters $N_1^j, \cdots N_n^j$, capturing a density function (see [6]) over such strategies. The expectation of $k$th action being adopted by $j$ is $\frac{N_k^j}{\sum N_i^j}$. Intuitively, each $N_k^j$ can be viewed as the number of times outcome $k$ (in this case action $k$) has been observed. The initial parameters adopted by agent $i$ represent its prior beliefs about agent $j$'s strategy. For simplicity, we assume that prior parameters are set uniformly (e.g., at 1), reflecting a uniform expectation for each of $j$'s actions (this is not a uniform prior over strategies, of course).

As in fictitious play, at each interaction agent $i$ should adopt a best response based on its current beliefs. Instead of a strategy profile, agent $i$ has a distribution over individual strategies and an induced distribution over profiles. However, the Dirichlet parameters permit the expectation of individual moves, and hence a best response, to be determined easily. When the interaction has ended, $i$ updates its beliefs by incrementing the parameters $N_k^j$ (where agent $j$ was observed to perform its $k$th action).[6]

---

[6] It is important to note that the agents are updating as if the sampled distribution were stationary, which it is not. Thus, convergence must be ensured by properties of best responses. We also note that the conclusions we draw below regarding the performance of Bayesian learning (versus fictitious play) are not intended to denigrate the Bayesian method. The fact is we are using priors about "initial" strategies as if they were beliefs about the fi-

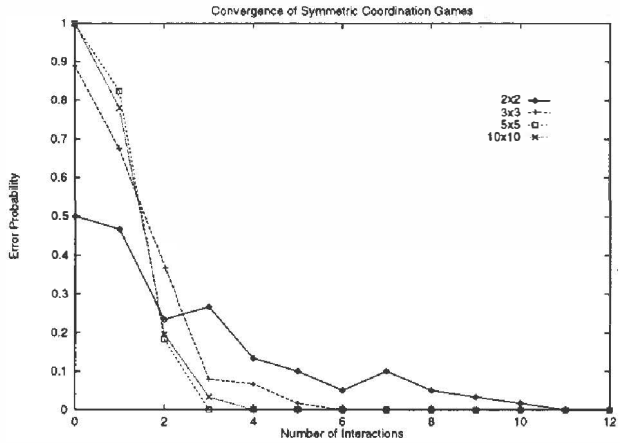

Figure 2: Convergence of Pure Coordination Games of Various Sizes. All results averaged over 30 trials.

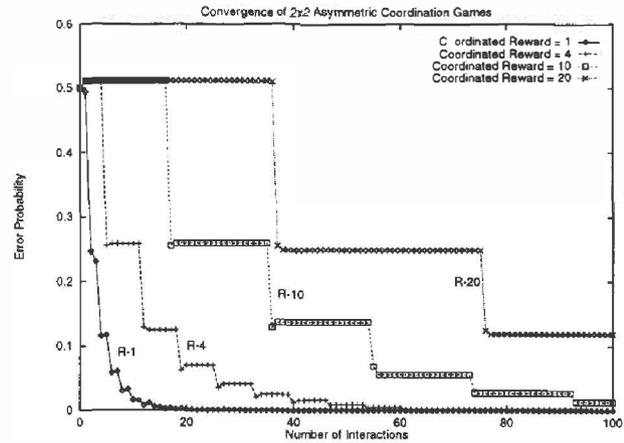

Figure 3: Convergence of $2 \times 2$ Asymmetric Games. Each game has a reward of 1 for one uncoordinated joint action, zero for the other, and the reward indicated (1, 4, 10, 20) for coordinated actions. All results averaged over 1000 trials.

In our example above (assuming observable actions), we might set the initial belief parameters of both agent $A$ and $B$ to $\langle 1, 1 \rangle$ (they each expect the other to go left or right with equal probability). Thus, they will each randomize between $l$ and $r$ uniformly. If the result of this randomization is coordinated (e.g., joint action $\langle l, l \rangle$), both update their distribution to be $\langle 2, 1 \rangle$. At the next interaction, both will adopt $l$ as a best response and reinforce the initial choice. It is easy to see that the OJA $\langle l, l \rangle$ is guaranteed to be selected forever. However, suppose the initial randomization results in the joint action $\langle l, r \rangle$. In this case, their updated beliefs will be different: $A$'s parameters $\langle 1, 2 \rangle$ indicate $B$ will again perform $r$, while $B$'s $\langle 2, 1 \rangle$ indicate the opposite. There is no chance of coordination at the next interaction: the action will be $\langle r, l \rangle$ (each switches actions). Their updated parameters will each be $\langle 2, 2 \rangle$ at this point and randomization can again take place providing another chance to coordinate at the third interaction.

It is not hard to see that, in this example, the agents have the opportunity to randomize at every second interaction, and the chance of coordination at each such round is 0.5. The probability that the agents fail to converge by round $k$ (i.e., $1/2^{\lfloor \frac{k}{2} \rfloor}$) therefore decreases exponentially with $k$. To illustrate, Figure 2 shows experimental results for this $2 \times 2$ games, as well as larger $n \times n$ *pure coordination games*.[7] The $x$-axis shows the number of times the game has been encountered, while the $y$-axis shows the average error probability—the chance an uncoordinated joint action is adopted using the agents's best response strategies at that point. In such pure coordination games, it is quite easy to see that convergence to an optimal joint action will be quite rapid. For instance, in the $10 \times 10$ game coordination is all but assured by the fourth play of the game.[8]

The rate of convergence can be adversely affected if the game is not symmetric. For example, consider the asymmetric $2 \times 2$ game given by:

|      | l(B) | r(B) |
|------|------|------|
| l(A) | 4    | 1    |
| r(A) | 0    | 4    |

Should the agents start with prior parameters $\langle 1, 1 \rangle$ representing their beliefs about the other's moves, then $A$'s initial best response is $l$, while $B$'s is $r$. The agents will not have the chance to coordinate their actions until they can randomize among their pure best responses—when $A$ assesses the probability of $r$ (for $B$) to be $\frac{4}{7}$ (or $B$ assigns probability $\frac{4}{7}$ to $l$). Given the integer nature of the updates, this can only happen at the sixth interaction, and every seventh interaction after that. Thus, the rate of convergence (while still exponential) is slowed linearly by a factor of seven. To illustrate the nature of these "plateaus", see Figure 3: values other than 4 (in the matrix above) are shown, along with the original $2 \times 2$ symmetric game.

**Proposition 1** *Let $G$ be a $2 \times 2$ coordination game, with $a$ denoting the utility of coordinated action, and $b, c$ denoting the utility of the two uncoordinated actions. Assuming uniform Dirichlet prior parameters $\langle 1, 1 \rangle$, the probability of failing to reach convergence at round $k$ is $1/(2^{\lfloor \frac{k}{g} \rfloor})$, where $g = \gcd((a - c), (a - c) + (a - d))$.*

---

nal "coordinated" strategies. It is remarkable that this misuse of Bayesian methodology works at all.

[7] In each game there are $n$ agents with $n$ actions. The set of moves is the same for all agents and they are rewarded with value $c$ if they each execute the same move, and are given a smaller value $d$ if they do not. Hence, there are $n$ OJAs.

[8] In fact, for larger values of $n$, faster convergence is due to the likelihood that the each randomization is more likely to produce a unique "most likely" (or majority) coordinated action.

Thus convergence is slowed linearly by the factor $g$. This can be extended to noninteger utilities in the obvious way; as long as the utilities are rational, convergence is guaranteed. We also note that nonuniform priors have little effect here, and that more heavily weighted priors do not preclude convergence, but can force a certain minimum number of encounters before coordination is possible.

## 4 Learning with Unobservable Stochastic Actions

The key difficulty with the models described above is the assumption that actions can be observed. As described at the outset, agents will typically be able to observe only the outcomes of these actions, and not the actions themselves. However, since the agents all know the game structure, the observations they make still provide evidence regarding the choices made by other agents. One simply needs to account for the inherent uncertainty in this information.

It is worth noting that, in general, there must be a sufficient number of observable states that can be used to distinguish (probabilistically) which joint actions have been executed for useful learning to take place. For instance, suppose we have simple matrix game where agents move to a good state or a bad state. If they can't observe the action chosen by others when moving to a bad state, then they can't tell which of the uncoordinated moves other agents did (i.e., very little information is available from which to learn). Our perspective is not so much that agents have a choice of actions that, correctly chosen, take them to a (single) good state (which is one interpretation of strategic form); rather they have a choice of possible good states, and their actions must be coordinated in the sense of agreeing on the state they "aim" for. (These are, of course, extreme points on a spectrum.)

### 4.1 Bayesian Best-Response Adapted

The Bayesian best-response model we described above can be adapted to the case of unobservable stochastic actions in a rather straightforward way. As before, we assume agents use Dirichlet distributions over the strategies of others to represent their beliefs. While belief parameters cannot be updated directly with observation of a particular action, agent $i$ can update its beliefs about $j$'s strategy by a simple application of Bayes rule. Agent $i$ first computes the probability that $j$ performed $a$ for any $a^j \in A_j$, given the observed state $s$ and its previous action $a^i$:

$$Pr(a[j] = a^j | a[i] = a^i, t) = \frac{Pr(t|a[j] = a^j, a[i] = a^i) Pr(a[j] = a^j)}{Pr(t|a[i] = a^i)}$$

Here $a[j]$ denotes $j$'s component of a joint action $a$. The prior probabilities are computed using agent $i$'s beliefs $Bel_i(k, a^k, s)$ for arbitrary agents $k$ and the joint transition probabilities. Agent $i$ then updates its distribution over $j$'s strategies using this "stochastic observation;" in particular $N_k^j$ is incremented by $Pr(a_k^j|t)$ (intuitively, by a "fractional" outcome).[9]

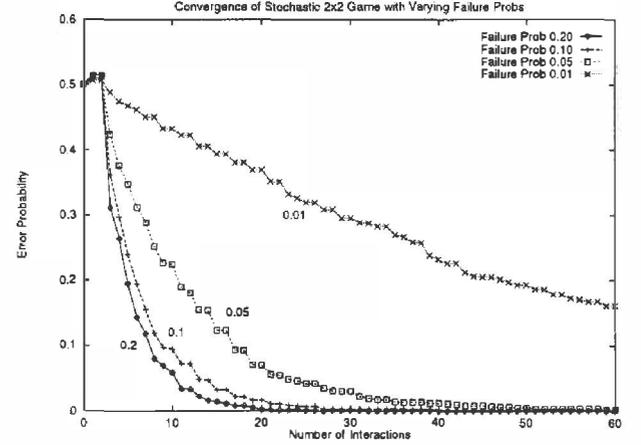

Figure 4: Convergence of $2 \times 2$ Stochastic, Unobservable Action Games. The probability of (individual) action failure $(0.01, 0.05, 0.1, 0.2)$ is shown. All results averaged over 1000 trials.

In the stochastic version of our example, let $A$ and $B$ adopt the initial parameters $\langle 1, 1 \rangle$. If the initial randomization results in coordinated joint action (e.g. $\langle l, l \rangle$), and the probable outcome $s_1$ results, coordination is assured forever. However, suppose the first joint action is $\langle l, r \rangle$ and it has its most likely outcome $s_4$. Then $A$'s belief parameters become $\langle 1.1, 1.9 \rangle$ and $B$'s $\langle 1.9, 1.1 \rangle$. The best response at the next interaction is $\langle r, l \rangle$, resulting in updated parameters $\langle 1.938, 2.061 \rangle$ for $A$ and $\langle 2.061, 1.938 \rangle$ for $B$ (assuming the expected outcome). Unlike the deterministic case, the agents will not be able to randomize or coordinate. In fact, given any sequence of "most likely outcomes," it is not hard to see that $A$ and $B$ will never coordinate, unless they do initially. Fortunately, this cycle of suboptimal joint actions can be broken by an unlikely outcome (i.e., if one of the actions "fails"). Experimental results for different failure probabilities in this $2 \times 2$ scenario are shown in Figure 4. These results illustrate the rather "paradoxical" fact that the less error prone (or more predictable) the available actions are, the slower the agents are to converge. Indeed, one can see that the stochastic actions play the role of "experimentation" for these agents, a technique used in game theory for agents to break out of suboptimal best response cycles.[10]

One way to enhance convergence is to have agents random-ize over $\varepsilon$-best responses [8]. This allows agents to randomize among actions that are close to being best responses given their current beliefs. In the example above, the beliefs of the agents "hover" around the point at which they will randomize—allowing $\varepsilon$-best responses gives the agents ample opportunity to break out of such cycles. This results in slightly better convergence in this example (see Figure 5).

---

[9] These fractional parameters correspond to the expectations of a weighted combination of integer-parameter Dirichlet distributions that result from standard update using the positive probability outcomes.

[10] Detailed, but straightforward, analysis of convergence using a Markov chain model is provided in a forthcoming technical report.

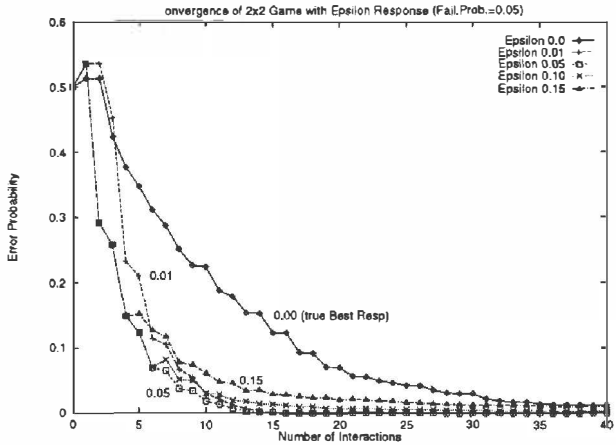

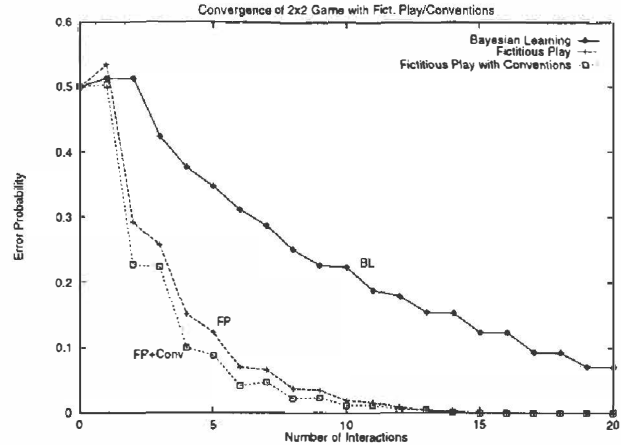

Figure 5: Convergence of 2 × 2 Stochastic, Unobservable Action Games With $\varepsilon$-Best Response. Various values of $\varepsilon$ (0.00, 0.01, 0.05, 0.1, 0.15) are shown. Action failure probability is 0.05. All results averaged over 1000 trials.

Figure 6: Convergence of 2 × 2 Games using Bayesian Learning, Stochastic Fictitious Play, and Fictitious Play with Conventions. Action failure probability is 0.05. All results averaged over 1000 trials.

### 4.2 Fictitious Play Adapted

Finally, we note that fictitious play can be adapted to the setting of unobservable stochastic actions with good success. Unlike the Bayesian model, we cannot rely on priors to estimate the probability a given action was performed. Instead we use *likelihood estimates* as a means of updating frequency counts in a way that accounts for the stochastic aspect of observations. When an outcome state $s$ is observed, each agent $i$ determines $Pr^a(s)$ for each joint action $a$ (this is just part of the agent's model). The relative likelihood of $a$ is $Pr^a(s)/\sum_b Pr^b(s)$, where $a$ and $b$ are restricted to range over joint actions such that $a[i] = a^i$, $b[i] = a^i$ (i.e., $i$ uses the knowledge of its *own* selected action $a^i$). Using these likelihoods, $i$ computes the likelihood that $j$ performed individual action $a^j$ to be

$$\frac{\sum\{Pr^a(s) : a[j] = a^j\}}{\sum Pr^a(s)}$$

(again, $a[i] = a^i$ is assumed). The likelihood estimates for each of these individual actions are used to update agent $i$'s frequency counts.

In our example, frequency parameters are updated by 0.9 or 0.1 for every possible outcome. This allows agents to randomize much more frequently, and is comparable to the action observable setting (for this example, not in general). Convergence for this version of fictitious play is compared to the Bayesian learning model for the 2 × 2 game (Figure 6), and a more complicated 3 × 3 game (Figure 7).

## 5 Conventions

Finally, we consider how true conventions might arise via learning. The problem with all of the models above, in the presence of stochastic actions, is that they cannot be said to converge to a true convention in the sense discussed in the introduction. By a conventional way of acting, we mean a fixed strategy that is applied to a given situation without requiring any special deliberation. The learning models described all have a chance of "popping out" of equilibrium (e.g., through a series of unlikely occurrences) though the probability of this generally decreases quickly over time. A more serious difficulty is that the agents must constantly update their beliefs and "reconsider" their choice of action (by recomputing possible best responses). This is certainly not in the spirit of conventions, or fixed rules of encounter, that one must simply *apply* to a given situation.

Intuitively, we would like agents to adopt some criterion that would allow them to identify that an optimal equilibrium has been reached, and that this realization is common knowledge. In this way, agents will eventually stop "thinking" about how to behave in a given state and simply act.[11] It is important to emphasize the role common knowledge plays here. If an agents use personal knowledge in the decision to jump to a conventional equilibrium (e.g., assessing the probability of a joint action using knowledge of its own action, or its personal prior), they risk adopting different conventions (and never reconsidering), perhaps *guaranteeing* suboptimal behavior from that point on.

---

[11] Adopting a convention in this sense does not mean that agents cannot adapt to changes in circumstance (e.g., the introduction of new agents). This would be reflected by the fact that the agents engage each other in a different state, for which the adopted convention does not apply.

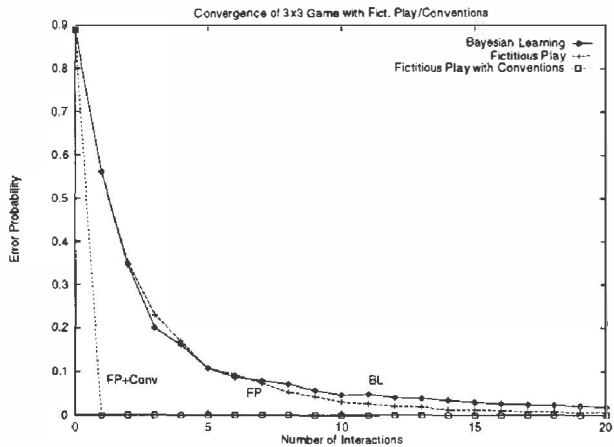

Figure 7: Convergence of $3 \times 3$ Games using Bayesian Learning, Stochastic Fictitious Play, and Fictitious Play with Conventions. Action failure probability is 0.05. All results averaged over 1000 trials.

We propose a model based on our stochastic extension of fictitious play and have agents compute the likelihood estimate of all OJAs (the "target" equilibria) given their current observation. If any OJA $a$ has a higher likelihood estimate than $b$, $b$ is "removed" from subsequent consideration. For instance, consider our $2 \times 2$ game. Suppose the agents end up in the state $\langle ll \rangle$; regardless of past behavior and what actual actions the agents performed, there is a unique OJA, $\langle l, l \rangle$, that has maximum likelihood. We notice that each agent can determine this independent of any personal information, and is aware that others have this ability as well—the OJAs with *maximum likelihood* are *common knowledge*. From this point on, the agents will perform $\langle l, l \rangle$, even if the initial action they performed was $\langle r, r \rangle$, and by chance it had this very unlikely outcome.

In a similar $3 \times 3$ game, once could imagine that moving to a certain state is most likely given two of the three OJAs (e.g., $\langle 1, 1, 1 \rangle$ and $\langle 2, 2, 2 \rangle$), but is less likely given the third (e.g., $\langle 3, 3, 3 \rangle$). When this state is observed, the agents will reject $\langle 3, 3, 3 \rangle$ as a potential equilibrium, will individually never consider performing action 3, and will never consider a joint action in which the other agents performed 3 to have positive probability at any future interaction: the rows in the matrix corresponding to the components of the rejected OJA will be effectively "deleted."

Formally, *conventions* are added to a learning model (such as fictitious play) as follows. At each interaction (say interaction $k$), each agent computes a likelihood estimate $LE(a)$ for each OJA $a$, given the observed outcome (we note that in fictitious play, these are computed for *all* joint actions, and will therefore be available anyway). The set $MLE(k)$ is the set of OJAs that have maximum likelihood. The game is then altered as follows: any action $a^i \in A_i$ that does not occur in any element of $MLE(k)$ is "deleted" from the game in the sense defined above. At interaction $k + 1$, coordination is attempted for this reduced game. If we are fortunate, the $MLE$ set will eventually be pared down to a singleton (or a set of OJAs with "interchangeable" components) and a convention will be reached that can never be dropped. This will not always be the case, of course, as we discuss below.

Conventions based on maximum likelihood estimates can be implemented "as is", with each agent randomly choosing actions and ruling out certain possibilities as warranted by $MLE$. However, this is unlikely to work well in scenarios with a sufficiently large number of outcomes, so that many states have zero probability of being reached by any OJA (e.g., imagine a $10 \times 10$ game where only a small fraction of the $10^{10}$ actions have positive probability of an "informative" outcome). In this case, the rate of convergence will be dictated by the probability of reaching an informative state given a random joint action (which can be tiny in a case like this). We actually want to use learning to bias agent responses in order to increase the probability of an OJA (or simply the chance of informative outcomes).

This is easily accomplished by grafting conventions onto the learning models described above, having agents maintain personal estimates of other agents's strategies and adopt best responses. Thus convergence to OJAs will occur even if $MLE$ does not prune actions. In an extended fictitious play model, this is straightforward. The only complication lies in the deletion of individual actions from the game: each agent $i$ must be sure that, in future updating and computation of best responses, the estimated frequencies of the actions deleted for agent $j$ are ignored. The relative frequencies of the *remaining* actions form the basis of best-response considerations at subsequent interactions.

Convergence for fictitious play with conventions is compared to straightforward fictitious play for our standard $2 \times 2$ game in Figure 6, and a more complicated $3 \times 3$ game in Figure 7. In both cases convergence is enhanced, remarkably so in the $3 \times 3$ case, where coordination is guaranteed after one interaction. Of course, this is an artifact of the game—each outcome state has a unique OJA with maximum likelihood.[12] While convergence is enhanced, we note that a more important function of conventions is their role in the eventual elimination of the computational burden associated with ongoing computation of best responses.

---

[12]Informally, this game has six outcome states, three "good" and three "bad". Each good state corresponds to an OJA in the sense that the OJA likely leads to that state. If only two of the three individual actions are the same, there is a small chance of moving to the good state corresponding to the majority action, and so on. The game was actually designed to prevent ordinary fictitious play from converging too quickly!

We note that conventions will not generally lead to a unique choice of OJA. For example, in a game with three OJAs, where two of them lead to the same outcomes with the same probabilities, nothing can distinguish the two from the point of view of likelihood. In other words, each action outcome accords the same likelihood to these two actions. In this case, the learning component will choose one of the two actions; but while conventional deliberations may rule out the third, they must leave open the possibility that either of the remaining two actions could be performed. In such a case, conventions cannot be used to prevent agents from continuing to update their beliefs.[13] However, in a case like this conventions still play a role in restricting attention in learning to particular possibilities, even if they cannot choose a unique equilibrium. The analysis of conventions and their effect on convergence is the subject of ongoing investigation and experimentation.

## 6   Concluding Remarks

We have studied several learning models from game theory, and their extension to coordination problems with unobservable actions. As we have seen, a number of different problem features, such asymmetries in utility and failure probabilities can have surprising effects of convergence to a coordinated equilibrium. We have also proposed the use of conventions as a means to restrict attention to par icular equilibria, in some cases allowing eventual relief from having to "think about" what action to perform.

The experimental results are not conclusive; rather they are merely suggestive of interesting models for coordination learning, models that require further exploration. However, some of these directions appear promising. In addition, the interaction of these methods in true sequential decision problems consisting of a wide variety of related state games is of considerable interest [2]. In this setting, we are ultimately interested in the generalization of learned conventions across similar state games, exploiting structured (Bayes net) representations of games and utility functions, as in [3]. Finally, generalizations of this model, especially those where only *partial* common knowledge of the game structure is assumed, will be required to make the effort more robust and realistic. This will require the use of ideas from reinforcement learning and learning models of dynamical systems.

**Acknowledgements:** This work benefited greatly from discussions with David Poole and Moshe Tennenholtz. This research was supported by NSERC Research Grant OGP0121843.

---

[13] In principle, one can detect this fact by analyzing the legal actions remaining at any point in the game and seeing if they can be distinguished by likelihood estimates. If the agents ever reach the point where (say, in this example) the two actions can never be distinguished, they can cease computing likelihood estimates, since the impossibility of reaching a convention has been detected.